\documentclass[a4paper,fleqn,usenatbib]{mnras}
\usepackage{newtxtext,newtxmath}

\usepackage[T1]{fontenc}
\usepackage{ae,aecompl}

\usepackage{booktabs}

\usepackage{graphicx}	
\usepackage{amsmath}	
\usepackage{amssymb}
\usepackage{subfigure}
\usepackage{natbib}
\usepackage[utf8]{inputenc} 
\usepackage{natbib}
\usepackage[usenames,dvipsnames]{xcolor}
\usepackage{cite}
\usepackage{float}
\usepackage{color}
\usepackage{hyperref}
\usepackage{multirow}
\usepackage{colortbl}
\definecolor{kugray5}{RGB}{224,224,224}

\title[CO Luminosity-FWHM Correlation]{CO luminosity-FWHM correlation of low- and high-redshift galaxies and its possible cosmological utilization}

\author[Yi-Han Wu et al.]{
Yi-Han Wu,$^{1}$\thanks{E-mail: s101022808@m101.nthu.edu.tw}
Tomotsugu Goto,$^{1}$
Ece Kilerci-Eser,$^{1, 2}$
Tetsuya Hashimoto,$^{1}$
Seong-Jin Kim,$^{1}$
\newauthor{\,\,Chia-Ying Chiang,$^{1}$ and Ting-Chi Huang$^{1}$}
\\
$^{1}$Institute of Astronomy, National Tsing Hua University, No. 101, Section 2 , Kuang-Fu Road, Hsinchu City 30013, Taiwan\\
$^{2}$Istanbul University, Science Faculty, Department of Astronomy and Space Sciences, Beyazıt, 34119, Istanbul, Turkey\\
}

\pubyear{2018}
\hypersetup{draft}
\begin{document}
\label{firstpage}
\pagerange{\pageref{firstpage}--\pageref{lastpage}}
\maketitle

\begin{abstract}
A linear correlation has been proposed between the CO luminosity\,($\rm{L}^{\prime}_{\rm{CO}}$) and full-width at half maximum\,(FWHM) for high-redshift\,($z > 1$) submillimeter galaxies. However, the controversy concerning the $\rm{L}^{\prime}_{\rm{CO}}$-FWHM correlation seems to have been caused by the use of heterogeneous samples (e.g., different transition lines) and/or data with large measurement uncertainties. In order to avoid the uncertainty caused by using different rotational transitions, in this work we make an extensive effort to select only CO($J=1-0$) data from the literature. We separate these wide-ranging redshift data into two samples : the low-redshift\,($z < 1$) and high-redshift\,($z > 1$) samples. The samples are corrected for lensing magnification factors if gravitational-lensing effects appeared in the observations. The correlation analysis shows that there exists significant $\rm{L}^{\prime}_{\rm{CO}}$-FWHM correlations for both the low-redshift and high-redshift samples. A comparison of the low- and high-redshift $\rm{L}^{\prime}_{\rm{CO}}$-FWHM correlations does not show strong evolution with redshift. Assuming that there is no evolution, we can use this relation to determine the model-independent distances of high-redshift galaxies. We then constrain cosmological models with the calibrated high-redshift CO data and the sample of Type Ia supernovae in the Union 2.1 compilation. In the constraint for $w$CDM with our samples, the derived values are $w_{0}=-1.02\pm{0.17}$,\,\,$\Omega_{m0}=0.30\pm{0.02}$, and\,\,$\rm{H}_{0}=70.00\pm{0.60}\,\,\rm{km\,s^{-1}\,Mpc^{-1}}$.
\end{abstract}


\section{Introduction}
The accelerating expansion of the Universe was revealed through the observations of Type Ia Supernovae\,(SNe Ia) in late 1990s\,\citep{1998AJ....116.1009R, 1999ApJ...517..565P}. Those distant SNe Ia are fainter than expected in a matter-dominated Universe, implying that dark energy drives the accelerating expansion. However, the current observations of SNe Ia are limited up to around $z \simeq 2$\,\citep{2013ApJ...768..166J} and the number of SNe Ia events decreases toward higher redshifts\,\citep{2011MNRAS.417..916G}. It may take a few Gyrs for a white dwarf binary to reach Chandrasekhar mass to explode. Due to these reasons, it will be difficult to investigate the variable nature of dark energy if we rely on SNe Ia only.

Thanks to improved sensitivity in infrared and radio observations during the past twenty years, many high-redshift\,($z > 1$) sub-millimeter galaxies\,(SMGs) and quasars\,(also known as Quasi-Stellar Objects, QSOs) have been observed\,\citep{2013ARA&A..51..105C, 2013Natur.496..329R}. The state-of-the-art observations can reach up to $z\simeq 6$\,\citep{2013ApJ...773...44W} through the observations of atomic and molecular lines from distant galaxies. For this reason, it is becoming possible to investigate the evolution of dark energy beyond $z > 1$ where large variations of dark energy are expected \,\citep{2014MNRAS.441.3454K}. 

A correlation between the carbon monoxide\,(hereafter CO) luminosity and the CO linewidth, full width at half maximum\,(FWHM), for high-$z$ galaxies has been proposed\,\citep{2012ApJ...752..152H, 2013MNRAS.429.3047B, 2015A&A...579A..17G}. This empirical power-law relation was applied to determine the magnification factors\,($\mu$) of gravitational-lensed galaxies\,\citep{2012ApJ...752..152H}. Besides, as proposed by \citet{2015A&A...579A..17G}, the $\rm{L}^{\prime}_{\rm{CO}}$-FWHM correlation may be possibly used as a cosmic distance ladder toward high-redshift\,($z > 1$) galaxies once we assure that the relation is redshift-independent. However, \citet{2016MNRAS.457.4406A} and\,\citet{2016ApJ...827...18S} did not find a significant correlation between the CO luminosity and FWHM. As found out the reason, these studies used highly heterogeneous samples with different rotational transitions of CO, and therefore could not see a clear correlation.   

To resolve those problems, in this work we focus only on the $J=1-0$ rotational transition of CO, and compile a large sample. Our compiled sample is divided into two groups according to the redshift ranges: low redshift ($z < 1$) and high redshift ($z > 1$) samples. In this work, we examine the existence of the $\rm{L}^{\prime}_{\rm{CO(1-0)}}$-FWHM relation for the low-$z$ and high-$z$ samples separately. We then investigate the redshift dependence of the $\rm{L}^{\prime}_{\rm{CO(1-0)}}$-FWHM relation. We also investigate the possible application of this relation through performing constraints of cosmological parameters.
	
In Section 2, we present the details of our data compilation and selection. Section 3 presents the methods in our analysis. Section 4 shows the results. Section 5 and 6 present the discussions and conclusions for the obtained results.  
  
\section{Data}
We searched for all the previous references for the galaxies taken by CO($J=1-0$) observations. Our compiled galaxies are divided into the low-$z$\,($z < 1$) and high-$z$\,($z > 1$) samples. In the following we describe the process of our compilation. 
\subsection{Low-$z$ Sample}
\label{low-z}
\citet{1995ApJS...98..219Y} used a $14$-meter telescope at the Five College Radio Astronomy Observatory\,(FCRAO) to carry out the CO($J=1-0$) observations for $300$ nearby galaxies\,($0.00 < z < 0.07$). $107$ of the $300$ galaxies were single-position observations since the angular sizes of the $107$ galaxies were generally smaller than $3^{\prime}$, and the rest was observed in multiple-position observation because the angular size was larger than $3^{\prime}$.

First we selected $107$ galaxies that were covered by single-position observations from the FCRAO catalogue. These single-position observations allow us to characterize the dynamical properties of the entire galaxies easily. Then, we selected $57$ of the $107$ single-position galaxies in which CO emission lines were detected and their redshifts\,($z$), velocity-integrated intensities\,($I_{\rm{CO}}$, $\rm{K\,km\,s^{-1}}$), and FWHMs\,($\rm{km\,s^{-1}}$) were all given in the literature. The global CO velocity-integrated fluxes\,($\rm{Jy\,km\,s^{-1}}$), which had been measured by fitting the model that closely resembled the observed CO spatial distributions, were also provided for these $57$ galaxies. We excluded one galaxy, $\rm{DDO69}$, from the $57$-galaxy sample since its FWHM\,($25 \rm{km\,s^{-1}}$) was particularly small and comparable to the velocity resolution of the spectrum\,($15 \rm{km\,s^{-1}}$) in the FCRAO observation. 

Three galaxies\,(IC883, Mrk231, and Mrk273) among the FCRAO sample had also been observed in CO\,($J=1-0$) with Institute de Radioastronomie Millim$\acute{\mathrm{e}}$trique\,(IRAM) telescope and their observations were given in \citet{1997ApJ...478..144S}. In the IRAM catalogue\,\citep{1997ApJ...478..144S}, IC883 was identified as Arp193. The IRAM catalogue listed redshifts (z), velocity-integrated intensities\,($I_{\rm{CO}}$, $\rm{K\,km\,s^{-1}}$), and FWHMs\,($\rm{km\,s^{-1}}$) of IC883 and Mrk273. New observed data were also published for Mrk231 in the IRAM catalogue. Therefore, we decided to adopt the more recent data for these three galaxies. IZw1 and $09320+6134$ were also selected from the IRAM catalogue and added into our low-$z$ sample since we considered their measured redshifts are closer and more compatible to those of the other selected FCRAO galaxies, and these two galaxies were identified as single galaxies morphologically as well. Here, there are $53$ single-position galaxies from the FCRAO catalogue and $5$ galaxies from the IRAM catalogue.

For the $53$ single-position FCRAO galaxies, the uncertainties of their FWHM measurements were not given in the literature. Referred to \citet{1982ApJ...258..467Y}, we measured the line widths of the $53$ FCRAO galaxies by dividing the velocity-integrated intensity\,($I$, $\rm{K\,km\,s^{-1}}$) with the peak antenna temperature\,($A$, milliKelvin\,$\rm{mK}$), whose $1\sigma$ uncertainties, $\sigma_{I}$ and $\sigma_{A}$, were also provided in the literature. By checking the positive linear correlation between the FWHMs given in literature and the measured line widths above, we proposed derived FWHMs\,($\rm{FWHM_{deri.}} \propto \frac{I}{A}$) should be identical with the literature ones and suggested 
\begin{equation}
\label{eq:FWHM same}
\frac{\sigma_{\rm{FWHM}}}{\rm{FWHM}} = \frac{\sigma_{\rm{FWHM_{deri.}}}}{\rm{FWHM_{deri.}}} = \sqrt{(\frac{\sigma_I}{I})^{2}+(\frac{\sigma_{A}}{A})^2}.
\end{equation}  
$\sigma_{\rm{FWHM}}$ and $\sigma_{\rm{FWHM_{deri.}}}$ are the $1\sigma$ uncertainties of the literature FWHM and $\rm{FWHM_{deri.}}$. Based on Eq.~\ref{eq:FWHM same}, we can assign the uncertainty to the literature FWHM of the selected FCRAO sample galaxy. For IC883, Mrk231, Mrk273, IZw1, and $09320+613$ that selected from the IRAM catalogue, the uncertainties of their FWHM measurements were given.  

\begin{figure}
\includegraphics[scale=0.33]{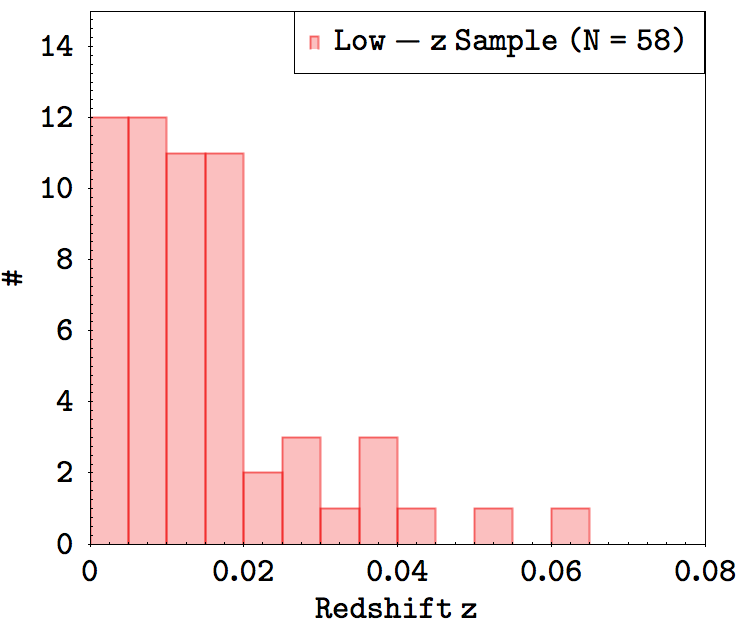}
\caption{The redshift distribution of the low-$z$ sample\,(listed in Table~\ref{tab:tabb1}) after the selection. The total number is $58$.}
\label{fig:low_z_z_distribution}
\end{figure}

Finally, $58$ galaxies were selected into our low-$z$\,($z < 1$) sample. Fig.~\ref{fig:low_z_z_distribution} presents the redshift distribution of the low-$z$ sample. Table ~\ref{tab:tabb1} in Appendix~\ref{S:Appendixgalaxy} lists the references and measurements of the selected galaxies in the low-$z$ sample.   

\subsection{High-$z$ Sample} 
The CO observations of high-$z$\,($z > 1$) galaxies published during the past twenty years were detected in multiple CO transitions. To avoid the uncertainty caused by utilizing different CO transitions and to perform a fair comparison with the low-$z$ sample, in our compilation, we focused only on the CO data in $J=1-0$ transition. 

Moreover, for those gravitational-lensed high-$z$ galaxies in the literature, we chose those whose magnification factors\,($\mu$) were available with measurement errors to reduce the uncertainty in the subsequent regression analysis.

\begin{figure}	
\includegraphics[scale=0.33]{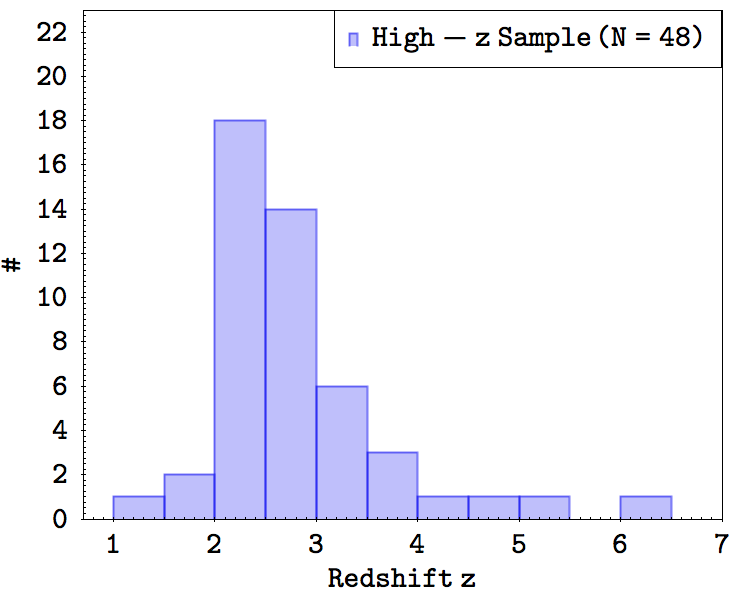}
\caption{The redshift distribution of the high-$z$ sample ($1.0 < z < 6.5$) in Table~\ref{tab:tabb2}.}
\label{fig:high_z_z_distribution}
\end{figure}

In total, $30$ SMGs, $10$ QSOs, four star-forming radio galaxies\,(SFRGs), two radio galaxies\,(RGs), and two Lyman-break galaxies\,(LBGs) were selected with secure redshift\,$z$, FWHM\,($\rm{km\,s^{-1}}$), velocity-integrated flux ($\rm{Jy\,km\,s^{-1}}$), and magnification factor\,($\mu$), whose their $1\sigma$ uncertainties were available in the literature. Fig.~\ref{fig:high_z_z_distribution} presents the redshift distribution of the high-$z$ samples. Table~\ref{tab:tabb2} in Appendix~\ref{S:Appendixgalaxy} lists the references and measurements of the selected galaxies in the high-$z$ sample.

\section{Analysis}
\subsection{Inclination Correction}
The measured CO FWHM\,($W_{\rm{CO}}$) is a projected quantity strongly relevant to the inclination of a galaxy, and they can be simply corrected into an intrinsic one by following relation 
\begin{equation}
\label{eq:FWHM correction}
W^{\rm{corr.}}_{\rm{CO}}=\frac{W_{\rm{CO}}}{\rm{sin}(\it{i})},
\end{equation}  
where $W^{\rm{corr.}}_{\rm{CO}}$ and $W_{\rm{CO}}$ are the corrected and measured FWHMs and $i$ is the inclination of a galaxy in degree.

Most of the inclinations of our low-$z$ sample galaxies are available in literature and other databases. But, it has been difficult to define the inclinations of the high-$z$ sample galaxies due to the limitation on spatial resolution in the current observations. Hence, to treat the low-$z$ and high-$z$ samples in the same way for subsequent analysis, we did not consider any inclination correction for all samples. 

\subsection{CO Luminosity and $\mu$ Correction}
\label{CO Luminosity and mu Correction}
We evaluated CO luminosities to our selected galaxy samples with the following equation\citep{1997ApJ...478..144S, 2011ApJ...734L..25E, 2012A&A...548A..22M, 2013ARA&A..51..105C}.
\begin{equation}
\label{eq: Luminosity equation}
L^{\prime}_{\rm{CO}}=\frac{3.25\times10^7\,I_{\rm{CO}}\,D_{\rm{L}}^2}{{\nu}^2_{0}(1+z)}\,\,\rm{K\,\,km\,\,s^{-1}\,\,pc^{2}}. 
\end{equation}
$I_{\rm{CO}}$\,($\rm{Jy\,km\,s^{-1}}$) and $D_{\rm{L}}$\,($\rm{Mpc}$) are the CO velocity-integrated flux and the luminosity distance, respectively. ${\nu}_{0}$ is the rest-frame frequency of CO($J=1-0$) emission in the unit of $\rm{GHz}$ and $z$ is the redshift of a galaxy. For those FCRAO galaxies in the low-$z$ sample, we used the global fluxes, which was measured in $\rm{Jy\,km\,s^{-1}}$, to derive their values of $L^{\prime}_{\rm{CO}}$. For the IRAM galaxies, however, the original values of the $I_{\rm{CO}}$ were measured in $\rm{K\,km\,s^{-1}}$. Thus, we converted those measurements into $\rm{Jy\,km\,s^{-1}}$ by multiplying $4.5 \rm{Jy\,K^{-1}}$\,\citep{1997ApJ...478..144S}. For the high-$z$ sample, the $I_{\rm{CO}}$ was all measured in $\rm{Jy\,km\,s^{-1}}$. The values of the $D_{\rm{L}}$ for our selected galaxies were derived with their given redshifts based on the concordance flat $\Lambda$CDM model. The parameters of the model are : $\rm{H}_{0}=73.8\,\rm{km\,\,s^{-1}\,\,Mpc^{-1}}$\,\citep{2011ApJ...730..119R}, $\Omega_{m0}=0.295$, $\Omega_{{\Lambda}0}=0.705$\,\citep{2012ApJ...746...85S}. 

We used Eq.~\ref{eq: Luminosity equation} to measure the CO luminosity, but a correction is required if a galaxy is gravitationally-lensed. We transferred the measured CO luminosity in Eq.~\ref{eq: Luminosity equation} into the intrinsic one by using Eq.~\ref{eq:L_conversion}\,\citep{2011ApJ...730..108R,2014MNRAS.440.1999N}. 
\begin{equation}
\label{eq:L_conversion}
L^{\rm{intris.}}_{\rm{CO}}=\frac{L^{\prime}_{\rm{CO}}}{\mu}.
\end{equation}
$L^{\rm{intris.}}_{\rm{CO}}$ and $L^{\prime}_{\rm{CO}}$ are the intrinsic and measured CO\,($J=1-0$) luminosities. $\mu$ is a dimensionless magnification factor and $\mu > 1$ denotes that a galaxy is observed with a gravitationally-lensed effect, while $\mu =1$ indicates that a galaxy is observed without any gravitationally-lensed effect. For our selected CO($J=1-0$) galaxies in the low-$z$ sample, the lensing effect has not been detected among them so the $\mu$ correction was not applied.

\subsection{The Linear Regression}
We first investigated the possibility of redshift independence for the correlations between the intrinsic CO luminosity\,($L^{\rm{intrins.}}_{\rm{CO}}$) and uncorrected FWHM\,($W^{\rm{uncorr.}}_{\rm{CO}}$) in the low-$z$ and high-$z$ samples. Statistically, the slope of a linear regression line is sensitive to the measurement errors of the independent variables, and when errors of one variable are smaller than the other, one can obtain more robust results by treating the one variable with smaller errors as the independent variable, and perform regression on the other with larger errors. In our case, we noted that the measurement errors of the $L^{\rm{intrins.}}_{\rm{CO}}$ are generally smaller than those of the $W^{\rm{uncorr.}}_{\rm{CO}}$ by comparing the sizes of errors of luminosities and FWHMs in the normalized space, i.e. normalized by the r.m.s of the data distributions. Hence, we decided to perform the regression line using the luminosity as the independent variable. 

We adopt a power-law parametrization in a linear logarithmic forms, as shown in Eq~\ref{eq:linear relation}.
\begin{equation}
\label{eq:linear relation}
\rm{log}_{10}(W^{\rm{uncorr.}}_{\rm{CO}}) - A=\alpha + \beta \,(log_{10}(L^{\rm{intrins.}}_{\rm{CO}}) - B),
\end{equation} 
where $\alpha$ and $\beta$ are the intercept and slope, and $A$ and $B$ the pivot points in the axes of $\rm{log}_{10}(W^{\rm{uncorr.}}_{\rm{CO}})$ and $\rm{log}_{10}(L^{\rm{intrins.}}_{\rm{CO}})$. We assigned the median values of the whole samples as the values of the pivot points, with $B=9.81$ for $\rm{log}_{10}(L^{\rm{intrins.}}_{\rm{CO}})$ and $A=2.45$ for $\rm{log}_{10}(W^{\rm{uncorr.}}_{\rm{CO}})$. We took the measurement errors of the $\rm{log}_{10}(L^{\rm{intrins.}}_{\rm{CO}})$ and $\rm{log}_{10}(W^{\rm{uncorr.}}_{\rm{CO}})$ into account as using the code \texttt{LINMIX{\_}ERR} in \texttt{IDL}\,\citep{2007ApJ...665.1489K} in the fitting procedure based on Eq.~\ref{eq:linear relation}. The code also derives the intrinsic scatter,\,$\epsilon_{i}$, for each fitting process, implying a probability of taking the $i$-th y data at fixed $i$-th x data respect to a fitting line. 
\section{Results}
\label{154927_8Dec17}
\begin{figure*}
\centering
\includegraphics[scale=0.50]{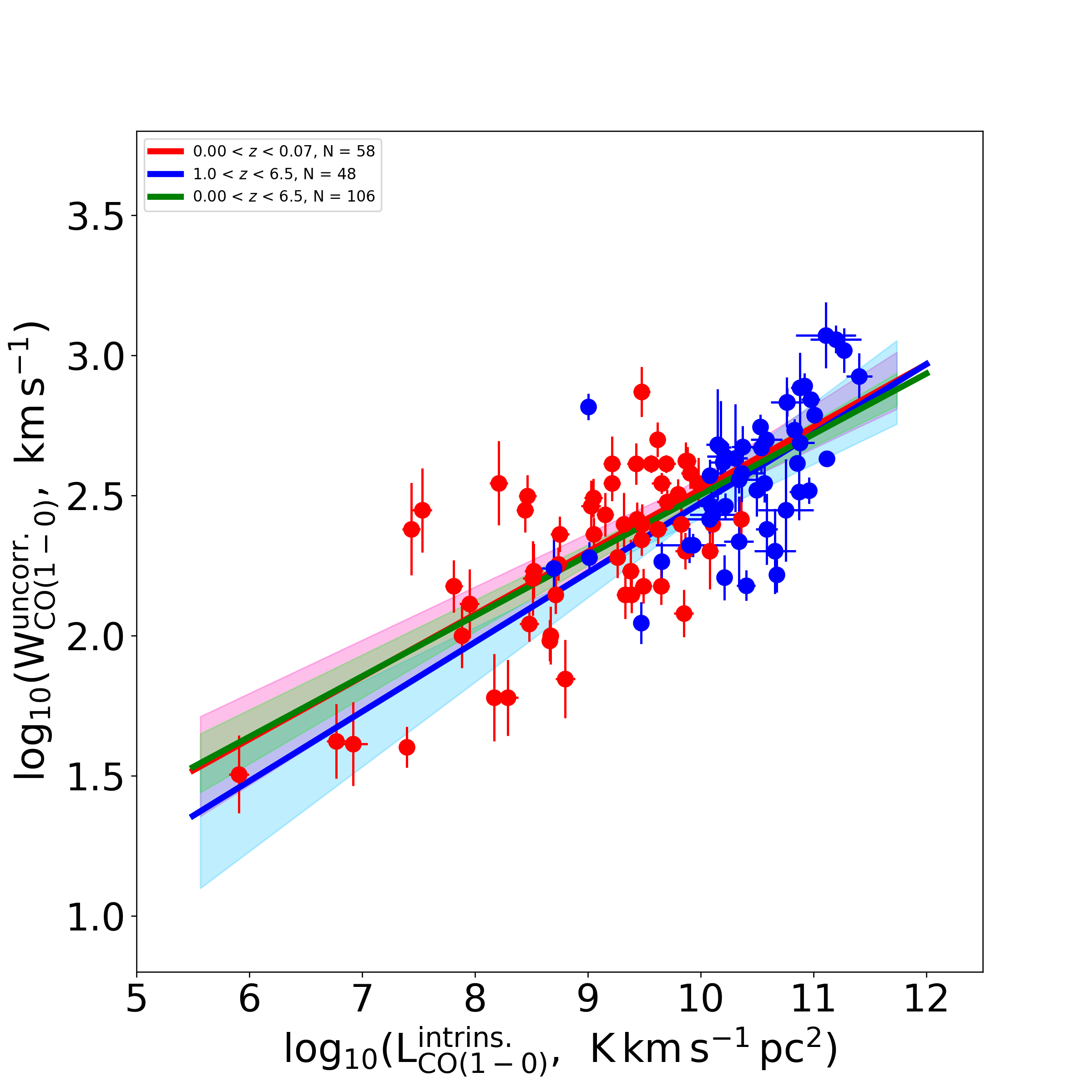}
\caption{The uncorrected FWHMs\,($W^{\rm{uncorr.}}_{\rm{CO(1-0)}}$) versus the intrinsic CO($J=1-0$) luminosities\,($L^{\rm{intrins.}}_{\rm{CO(1-0)}}$) for the low-$z$, high-$z$ in logarithmic plane. The red and blue dots denote the low-$z$ and high-$z$ galaxies. The red, blue, and green lines indicate the best-fit lines of the low-$z$, high-$z$, and whole samples\,(low-$z$ + high-$z$), with corresponding red, blue, and green shaded regions around the best-fit lines. These shaded regions present the $1-\sigma$ uncertainties around the best-fit lines. Table~\ref{tab:the linear regressions} presents the parameters of the best-fit lines in this diagram.}
\label{fig:correlation L and F}
\end{figure*}
\begin{table*}
\centering
\caption{The best-fit results from the linear regressions for the low-$z$ ($0.00 < z < 0.07$), high-$z$ ($1.0 < z < 6.5$), and whole\,($0.0 < z < 6.5$) samples in Fig.~\ref{fig:correlation L and F}. The values that are behind the plus-minus signs are $1-\sigma$ uncertainties. $N$ is the number of galaxies in each sample. The $i$-th intrinsic scatter\,(IS),\,$\epsilon_{i}$, implies the probability of taking the $i$-th y variable at the fixed x variable. Spearman's coefficient\,(SC) demonstrates the degree of the power-law correlation between the two variables: $L^{\rm{intrins.}}_{\rm{CO(1-0)}}$ and $W^{\rm{uncorr.}}_{\rm{CO(1-0)}}$. The $p$ value presents the significance of a correlation and $p < 0.05$ can be considered as a significant correlation. The r.m.s. means the root-mean-square value of the predicted and observed y data respect to a fitting line. The $\rho$ or Cov.($\alpha$, $\beta$) is the covariance of the intercept and slope for the sample.}
\label{tab:the linear regressions}
\begin{tabular}{lllllllll}
\toprule[3pt]
Sample & $N$ &Intercept\,($\alpha$) &Slope\,($\beta$)&IS\,($\epsilon_{i}$) & SC & $p$ value & r.m.s\,($\rm{km\,s^{-1}}$) & $\rho=$Cov.($\alpha$, $\beta$)\\
\hline
&&&&&&&&\\
Low-$z$ & $58$ &$0.03\pm{0.04}$&$0.22\pm{0.03}$ & $0.20\pm{0.02}$ & $0.54$&$1.33{\times}10^{-5}$ & $118$ & $0.68$ \\ 
&&&&&&&&\\
High-$z$ & $48$ &$-0.02\pm{0.05}$&$0.25\pm{0.05}$ & $0.20\pm{0.03}$ & $0.61$&$4.16{\times}10^{-6}$ & $202$ & $-0.72$ \\ 
&&&&&&&&\\
Whole & $106$ &$0.02\pm{0.02}$&$0.22\pm{0.02}$ & $0.19\pm{0.02}$ & $0.66$&$1.14{\times}10^{-14}$ & $163$ & $0.15$ \\
\bottomrule[3pt]
\end{tabular}
\end{table*}

Fig.~\ref{fig:correlation L and F} presents the correlations between the uncorrected CO($J=1-0$) FWHMs\,($W^{\rm{uncorr.}}_{\rm{CO(1-0)}}$) and intrinsic CO($J=1-0$) luminosities\,($L^{\rm{intrins.}}_{\rm{CO(1-0)}}$) for the low-$z$\,(red dots), high-$z$\,(blue dots), and whole samples. The red, blue, and green lines indicate the regression lines for the low-$z$ and high-$z$ galaxies, and whole samples, with corresponding red, blue, and green shaded regions representing the $1-\sigma$ uncertainties to the best-fit lines. The best-fit results in Fig.~\ref{fig:correlation L and F} are listed in Table~\ref{tab:the linear regressions}. As noted in the table, the Spearman's coefficient of the correlation of the low-$z$ sample is $0.54$ with the corresponding $p$ value of $1.33{\times}10^{-5}$, manifesting that there is a strong and significant correlation between the $L^{\rm{intrins.}}_{\rm{CO(1-0)}}$ and $W^{\rm{uncorr.}}_{\rm{CO(1-0)}}$ since $p < 0.05$ can be considered as a significant correlation. Similarly, the high-$z$ galaxies manifest stronger and more significant correlations between the $L^{\rm{intrins.}}_{\rm{CO(1-0)}}$ and $W^{\rm{uncorr.}}_{\rm{CO(1-0)}}$ than those in the low-$z$ sample, with Spearman's coefficient of $0.61$ and the $p$ value of $4.16{\times}10^{-6}$. 

By comparing the best-fit parameters of the low-$z$ and high-$z$ correlations in Table~\ref{tab:the linear regressions}, we found that the slopes of the low-$z$ and high-$z$ correlations are consistent with each other within $1-\sigma$ uncertainties of their slopes and their intercepts agree with each other within a $1-\sigma$ level. The consistency was also reflected in Fig.~\ref{fig:correlation L and F}, which $1-\sigma$ uncertainty regions around the low-$z$ and high-$z$ relations largely overlap with each other. 

\section{Discussions}  
As found in Sec. \ref{154927_8Dec17}, there are significant correlations between the intrinsic CO luminosity\,($L^{\rm{intrins.}}_{\rm{CO(1-0)}}$) and uncorrected FWHM\,($W^{\rm{uncorr.}}_{\rm{CO(1-0)}}$) existing in the low-$z$ and high-$z$ sample galaxies. Moreover, through the comparison of the slopes and intercepts of the low-$z$ and high-$z$ $L^{\rm{intrins.}}_{\rm{CO(1-0)}}$-$W^{\rm{uncorr.}}_{\rm{CO(1-0)}}$ correlations, they do not show a significant sign of redshift evolution. Next, based on these statistical trends, we discuss how to use the $L^{\rm{intrins.}}_{\rm{CO(1-0)}}$-$W^{\rm{uncorr.}}_{\rm{CO(1-0)}}$ correlation in cosmology. 

\subsection{Possible Utilization in Cosmology}
\subsubsection{Estimation of the Cosmology-Independent Distance To the High-$z$ Sample}
\label{Estimation of the Cosmology-Independent Distance To the High-$z$ Sample}
If there is no evolution in redshift of the $L^{\rm{intrins.}}_{\rm{CO(1-0)}}$-$W^{\rm{uncorr.}}_{\rm{CO(1-0)}}$ relation, it implies that we can potentially utilize this empirical scaling relation as a tool to estimate the model-independent luminosities or cosmic distances toward high-redshift regions. 

In Sec.~\ref{154927_8Dec17}, those relations were constructed based on the concordance ${\Lambda}CDM$ model in a flat Universe. Here, for the purpose of measuring cosmic distances, we need model-independent distance calibrators to calibrate the low-$z$ relation. In the process of our distance calibration, the $23$ ($0.0152 < z < 0.07$) of the $58$ our low-$z$ sample galaxies were calibrated with the SNe Ia data of the Union 2.1 sample\,\citep{2012ApJ...746...85S}, which provided the observed distance modulus\,(DM) and the corresponding $1-\sigma$ errors\,($\sigma_{\rm{DM}}$) of the $580$ SNe Ia data. The other $17$ ($0.0009 < z < 0.015$) of the low-$z$ galaxies were calibrated with the Cepheid-calibrated galaxies, which were catalogued in \citet{2001ApJ...553...47F}\,\,(see Table~\ref{tab:Cepheid Calibrated Galaxies}).
 
\begin{table}
\centering
\caption{The $27$ Cepheid-calibrated nearby galaxies published in \citet{2001ApJ...553...47F}. The first column shows the ID of the galaxy. The second column shows the redshift\,($z$) from the NED database. The third and fourth columns are the distance modulus\,(DM) and the corresponding $1-\sigma$ error\,($\sigma_{\rm{DM}}$).}
\begin{tabular}{lccc} 
\hline
Galaxy ID & $z$ & DM & $\sigma_{\rm{DM}}$\\
\hline\hline
NGC2403  		&	$0.000445$		&	$27.48$	&	$0.24$\\
NGC300  			&	$0.00048$		&	$26.53$	&	$0.07$\\
NGC5457 			&	$0.000804$		&	$29.13$	&	$0.11$\\
IC4182 			&	$0.001071$		&	$28.28$	&	$0.06$\\
NGC5253 			&	$0.001358$		&	$27.56$	&	$0.14$\\
NGC4258 			&	$0.001494$		&	$29.44$	&	$0.07$\\
NGC4548 			&	$0.001621$		&	$30.88$	&	$0.05$\\
NGC2541 			&	$0.001828$		&	$30.25$	&	$0.05$\\
NGC925 			&	$0.001845$		&	$29.80$	&	$0.04$\\
NGC3198 			&	$0.002202$		&	$30.68$	&	$0.08$\\
NGC4414 			&	$0.002388$		&	$31.10$	&	$0.05$\\
NGC3627 			&	$0.002425$		&	$29.86$	&	$0.08$\\
NGC3621 			&	$0.002435$		&	$29.08$	&	$0.06$\\
NGC3319 			&	$0.002465$		&	$30.64$	&	$0.09$\\
NGC3351 			&	$0.002595$		&	$29.85$	&	$0.09$\\
NGC7331 			&	$0.002722$		&	$30.81$	&	$0.09$\\
NGC3368 			&	$0.002992$		&	$29.97$	&	$0.06$\\
NGC2090 			&	$0.003072$		&	$30.29$	&	$0.04$\\
NGC4639 			&	$0.003395$		&	$31.61$	&	$0.08$\\
NGC4725 			&	$0.004023$		&	$30.38$	&	$0.06$\\
NGC1425 			&	$0.005037$		&	$31.60$	&	$0.05$\\
NGC4321 			&	$0.005240$		&	$30.78$	&	$0.07$\\
NGC1365 			&	$0.005457$		&	$31.18$	&	$0.05$\\
NGC4496A 		&	$0.005771$		&	$30.81$	&	$0.03$\\
NGC4536 			&	$0.006031$		&	$30.80$	&	$0.04$\\
NGC1326A 		&	$0.006108$		&	$31.04$	&	$0.09$\\
NGC4535 			&	$0.006551$		&	$30.85$	&	$0.05$\\
\hline
\label{tab:Cepheid Calibrated Galaxies}
\end{tabular}
\end{table}
We still utilized the pivot values in Sec.~\ref{154927_8Dec17} and calibrated the $L^{\rm{intrins.}}_{\rm{CO(1-0)}}-W^{\rm{uncorr.}}_{\rm{CO(1-0)}}$ relation by accomplishing the distance calibration to the available $40$ low-$z$ sample galaxies. The calibrated relation is formed as below     
\begin{equation}
\label{eq:calibrated L-FWHM correlation}
\rm{log}_{10}(W^{\rm{uncorr.}}_{\rm{CO(1-0)}}) - A = \alpha + \beta (log_{10}(L^{\rm{intrins.}}_{CO(1-0)}) - B).
\end{equation} 
$A$ and $B$ are the pivot values of $2.45$ and $9.81$. The $\alpha$, $\beta$, and $\epsilon_{i}$ are $0.03$, $0.21$, and $0.21$. The $1-\sigma$ uncertainties of the $\alpha$ and $\beta$ are $0.05$ and $0.04$. The $\rho$ or Cov($\alpha$, $\beta$) value of this calibrated relation is $0.59$. The $\alpha$ and $\beta$ of this calibrated relation are still consistent with the ones of the model-dependent relation in the low-$z$ sample in Sec.~\ref{154927_8Dec17} although only $40$ of the $58$ low-$z$ sample galaxies were involved in the distance calibration. 

We are able to derive the model-independent DMs and corresponding values of $\sigma_{\rm{DM}}$ to the high-$z$ sample galaxies through the use of Eq.~\ref{eq:calibrated L-FWHM correlation}. Briefly, the values of $\sigma_{\rm{DM}}$ of the high-$z$ sample galaxies are computed by starting from the error and scatter of Eq.~\ref{eq:calibrated L-FWHM correlation}, and the uncertainties of the $W^{\rm{uncorr.}}_{\rm{CO(1-0)}}$ of the high-$z$ sample galaxies are converted into the uncertainties of $L^{\rm{intrins.}}_{\rm{CO(1-0)}}$ of the same galaxies. Finally, the luminosity uncertainties are propagated into the uncertainties of the DM. We present the detail process of the derivation of the model-independent DMs and corresponding values of the $\sigma_{\rm{DM}}$ of the high-$z$ sample galaxies below. 

\begin{figure}	
\includegraphics[scale=0.35]{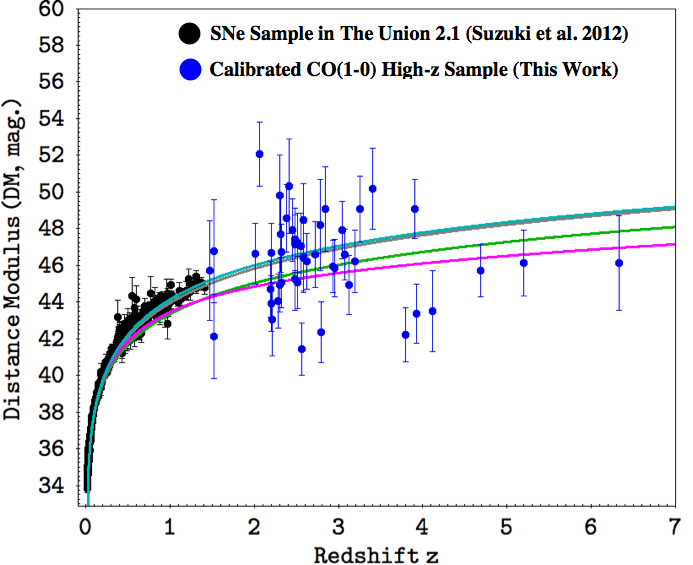}
\caption{The Hubble Diagram\,(HD) of the SNe Ia\,(the black dots) in the Union 2.1 and calibrated high-$z$ sample galaxies\,(the blue dots) in this work. The curves illustrate the cosmological models with different sets of parameters, which are shown in Table~\ref{tab:colored curves}.}
\label{fig:Hubble Diagram}
\end{figure}

\begin{figure}
\includegraphics[scale=0.36]{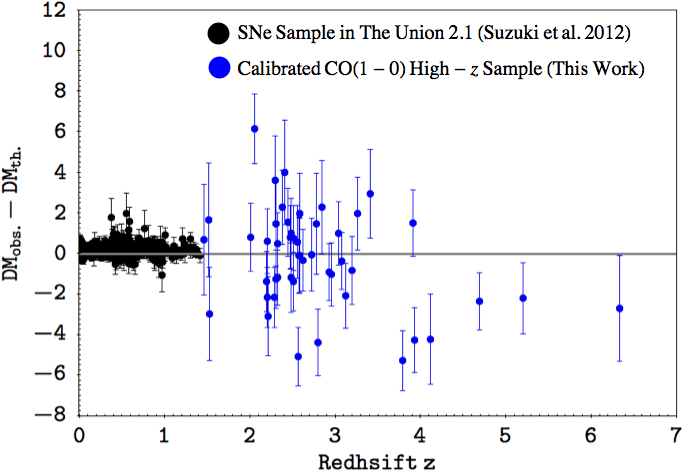}
\caption{The difference between the observational and theoretical DMs as function of redshift\,($z$). The observational DMs\,($\rm{DM}_{\rm{obs.}}$) of the SNe Ia are from Union 2.1. The $\rm{DM}_{\rm{obs.}}$ values of the high-$z$ sample galaxies are the calibrated values. The theoretical DMs\,($\rm{DM}_{\rm{th.}}$) at given $z$ are deduced based on a set of cosmological parameters\,($\Omega_{m0}$, $\Omega_{\Lambda 0}$, $w_{0}$, $w_{a}$)\,=\,($0.295$, $0.705$, $-1$, $0$) as in Fig.~\ref{fig:Hubble Diagram}. The black dots denote the data points of the SNe Ia and the blue ones the calibrated high-$z$ sample galaxies.}
\label{fig:muob-muth vs z}
\end{figure}

We estimate the model-independent luminosity by transferring Eq.~\ref{eq:calibrated L-FWHM correlation} into 
\begin{equation}
\label{eq: inverted L-FWHM correlation}
\rm{log}_{10}(L^{\rm{intrins.}}_{\rm{CO(1-0)}})=B + \frac{\rm{log}_{10}(W^{\rm{uncorr.}}_{\rm{CO(1-0)}})-A-\alpha}{\beta}.
\end{equation} 
The uncertainty of the logarithmic luminosity is defined based on the law of error propagation and shown term by term below 
\begin{align}
\label{eq: uncertainty log L}
&\sigma^{2}_{\rm{log}_{10}(L^{\rm{intrins.}}_{\rm{CO(1-0)}})}=\\\notag
&(\frac{\sigma_{\rm{log}_{10}(W^{\rm{uncorr.}}_{\rm{CO(1-0)}})}}{\beta})^{2}+(\frac{\sigma_{\alpha}}{\beta})^{2}+(\frac{ (\rm{log}_{10}(W^{\rm{uncorr.}}_{\rm{CO(1-0)}})-A-\alpha)\sigma_{\beta}}{{\beta}^{2}})^{2}\\\notag
&+\frac{ 2{\times}(\rm{log}_{10}(W^{\rm{uncorr.}}_{\rm{CO(1-0)}})-A-\alpha)\sigma_{{\alpha}{\beta}}}{{\beta}^{3}}+\epsilon_{i}.
\end{align}  
In Eq.~\ref{eq: uncertainty log L}, $\sigma_{\alpha}$ and $\sigma_{\beta}$ are the uncertainties of the intercept\,($\alpha$) and slope\,($\beta$) of the Eq.~\ref{eq: inverted L-FWHM correlation}. The uncertainty of the covariance term $\sigma_{{\alpha}{\beta}}={\rho}{\sigma_{\alpha}}{\sigma_{\beta}}$ is considered and its value can be obtained from the regression results of Eq~\ref{eq:calibrated L-FWHM correlation}. The intrinsic scatter $\epsilon_{i}$ of $0.21$ is also considered and added into Eq.~\ref{eq: uncertainty log L}. The model-independent luminosity distance\,($D_{\rm{L}}$, in $\rm{Mpc}$) can also be derived to the high-$z$ sample through the use of Eq.~\ref{eq: Luminosity equation} and Eq.~\ref{eq:L_conversion} if gravitationally magnification appears. The uncertainty of the $D_{\rm{L}}$ is derived from the equation
\begin{equation}
\label{eq: uncertainty DL 1st}
\sigma^{2}_{D_{\rm{L}}} = (\frac{D_{\rm{L}}}{2})^{2}[(\frac{\sigma_{F}}{F})^2 + (\frac{\sigma_{L^{\rm{intrins.}}_{\rm{CO(1-0)}}}}{L^{\rm{intrins.}}_{\rm{CO(1-0)}}})^2 + (\frac{\sigma_{\mu}}{\mu})^2 + (\frac{\sigma_{z}}{1+z})^2],
\end{equation}     
where we considered the measurement values and errors of the velocity-integrated flux\,($F$), estimated luminosity\,($L^{\rm{intrins.}}_{\rm{CO(1-0)}}$) in Eq.~\ref{eq: inverted L-FWHM correlation}, redshift\,($z$) and magnification\,($\mu$). We recall that $L^{\rm{intrins.}}_{\rm{CO(1-0)}}=10^{\rm{log}_{10}(L^{\rm{intrins.}}_{\rm{CO(1-0)}})}$ and $\sigma_{L^{\rm{intrins.}}_{\rm{CO(1-0)}}}=L^{\rm{intrins.}}_{\rm{CO(1-0)}}\rm{Ln}(10)\sigma_{\rm{log}_{10}(L^{\rm{intrins.}}_{\rm{CO(1-0)}})}$. Therefore, Eq.~\ref{eq: uncertainty DL 1st} can be also written as below
\begin{equation}
\label{eq: uncertainty DL 2nd}
\sigma^{2}_{D_{\rm{L}}} = (\frac{D_{\rm{L}}}{2})^{2}[(\frac{\sigma_{F}}{F})^2 + (\rm{Ln}(10)\sigma_{\rm{log}_{10}(L^{\rm{intrins.}}_{\rm{CO(1-0)}})})^2 + (\frac{\sigma_{\mu}}{\mu})^2 + (\frac{\sigma_{z}}{1+z})^2],
\end{equation} 
With the estimated model-independent $D_{\rm{L}}$ and the corresponding uncertainty in Eq.~\ref{eq: uncertainty DL 1st} or Eq.~\ref{eq: uncertainty DL 2nd}, the DM of the high-$z$ sample can be obtained with the equation
\begin{equation}
\label{eq : DM}
\rm{DM}=5\rm{log}_{10}(\frac{D_{\rm{L}}}{\rm{Mpc}})+25,
\end{equation}
with the uncertainty
\begin{equation}
\label{eq : uncertainty DM}
\sigma_{\rm{DM}}=\frac{5\sigma_{\rm{D}_{\rm{L}}}}{\rm{D}_{\rm{L}} \times Ln(10)}.
\end{equation}
\begin{table}
\centering
\caption{The parameters that are used to derive the model curves in Fig.~\ref{fig:Hubble Diagram}. All curves have common ${\sigma}_{8}$ and $H_{0}$ values, $0.8$ and $73.8\,\rm{km\,s^{-1}\,Mpc^{-1}}$.}
\begin{tabular}{ccccc} 
\hline
Color& $\Omega_{m0}$& $\Omega_{\Lambda 0}$& $w_{0}$& $w_{a}$\\
\hline
Green & $1$ & $0$ & $-1$ & $0$\\ 
Gray& $0.295$ & $0.705$ & $-1$ & $0$ \\
Pink & $0.295$ & $0.705$ & $-0.9$ & $2$\\
Cyan& $0.295$ & $0.705$ & $-0.9$ & $-2$\\ 
\hline
\label{tab:colored curves}
\end{tabular}
\end{table}
Fig.~\ref{fig:Hubble Diagram} displays the Hubble Diagram\,(HD), which plots the $\rm{DM}$ versus $z$, with the calibrated DMs of the high-$z$ sample galaxies and the SNe Ia of the Union2.1 in blue and black dots. The coloured curves in Fig.~\ref{fig:Hubble Diagram} illustrate the cosmological models, which we referred in\,\citet{1999astro.ph..5116H}, that are derived from different sets of parameters. These sets of parameters are presented in Table~\ref{tab:colored curves}. Fig.~\ref{fig:muob-muth vs z} illustrates the observational DM\,($\rm{DM}_{\rm{obs.}}$), with respect to the theoretical DM\,($\rm{DM}_{\rm{th.}}$), for the calibrated high-$z$ sample galaxies\,(the blue dots) and SNe\,(the black dots) within $0.015 < z < 7.0$. The $\rm{DM}_{\rm{th.}}$ is based on the concordance CDM whose the parameters are\,($\Omega_{m0}$, $\Omega_{\Lambda 0}$, $w_{0}$, $w_{a}$)\,=\,($0.295$, $0.705$, $-1$, $0$). The concordance CDM model is presented in the gray line in Fig.~\ref{fig:muob-muth vs z}.

\subsubsection{Constraints}
The cosmic expansion history or the evolution of dark energy can be traced via luminosity distances of distant bright sources. In this section, we demonstrate a possible utilization of the calibrated high-$z$ sample galaxies as a distance indicator to constrain cosmological parameters. 

Theoretically, the luminosity distance\,($\rm{D}_{\rm{L}}$) of a galaxy with a given redshift\,($z$) can be expressed by the following equation
\begin{equation}
\label{eq:luminosity distance}
\rm{D}_{\rm{L}}(\rm{Mpc})=\frac{c\,(1+z)}{H_{0}}\int^{z}_{0}\frac{dz^{'}}{E(z^{'},\textbf{p})},
\end{equation}   
where $c$ and $\rm{H}_{0}$ are the speed of light and the present-day Hubble constant. $E(z^{'},\textbf{p})$ is a function of redshift and a set of cosmological parameters\,($\textbf{p}$). Following\,\citet{2018arXiv180706209P}, we assumed a spatially-flat\,($\Omega_{k}=0$) Universe and $\Omega_{m}+\Omega_{\Lambda}=1$ in the following models for simplicity. However, we could also include the spatial curvature in the parametrization.  

In a $w$CDM model, the equation of state\,(EoS) of the dark energy\,($w$) is considered as constant as a present-day one\,($w=w_{0}=\rm{constant}$) and the $E(z,\textbf{p})$ is defined as
\begin{equation}
\label{eq:xWCDM E}
E(z|{\Omega_{m0}, w_{0}})=\sqrt{\Omega_{m0}(1+z)^{3}+(1-\Omega_{m0})(1+z)^{3(1+w_{0})}}.
\end{equation} 
$\Omega_{m0}$ is the present-day mass density. The $w$CDM model does not involve the evolution of dark energy. A CPL model\,\citep{2001IJMPD..10..213C, 2003PhRvL..90i1301L} has been proposed to consider the evolution of dark energy and the EoS of dark energy in this model is parametrized as
\begin{equation}
\label{eq:wz}
w=w_{0}+w_{a}(\frac{z}{1+z}),
\end{equation}
where $w_{a}$ is an evolutionary factor and 
\begin{align}
&E(z|{\Omega_{m0}, w_{0}, w_{a}})=\\
&\sqrt{\Omega_{m0}(1+z)^{3}+(1-\Omega_{m0})(1+z)^{3(1+w_{0}+w_{a})}\rm{exp}(\frac{-3w_{a}z}{1+z})}.\notag
\end{align}
We constrained the cosmological parameters for the $w$CDM and CPL models by finding the maximum value of the likelihood function $\mathcal{L} \propto \rm{exp}(\frac{-{\chi}^{2}}{2})$. The definition of $\chi^{2}$ is
\begin{equation}
\label{eq: chi square}
\chi^{2}=\sum^{N}_{i=1}(\frac{\rm{DM}^{\rm{obs.}}_{i}(z_{i})-\rm{DM}^{\rm{th.}}_{i}(z_{i})}{{\sigma}_{\rm{DM}^{\rm{obs.}}_{i}}} )^{2},
\end{equation}
where $\rm{DM}^{\rm{obs.}}_{i}(z_{i})$ and $\rm{DM}^{\rm{th.}}_{i}(z_{i})$ are the observational and theoretical DMs of the $i$-th source at given $i$-th redshift,\,$z_{i}$. The ${\sigma}_{\rm{DM}^{\rm{obs.}}_{i}}$ is the observational error of the $i$-th $\rm{DM}^{\rm{obs.}}_{i}$. 

The Union 2.1 provided the $\rm{DM}^{\rm{obs.}}_{i}$ and ${\sigma}_{\rm{DM}^{\rm{obs.}}_{i}}$ of the $580$ SNe Ia. The values of the $\rm{DM}^{\rm{obs.}}_{i}$ and ${\sigma}_{\rm{DM}^{\rm{obs.}}_{i}}$ of the calibrated high-$z$ sample galaxies were obtained from Eq.~\ref{eq : DM} and Eq.~\ref{eq : uncertainty DM}, following the process in Sec.~\ref{Estimation of the Cosmology-Independent Distance To the High-$z$ Sample}. The theoretical DM of an object at the given $z$ was computing by using Eq.~\ref{eq : DM} and Eq.~\ref{eq:luminosity distance}. 

To examine the effects of the improvement on the constraints in the $w$CDM and CPL models as adding the calibrated high-$z$ sample, two datasets, SNe\,(SNe Ia only) and SNe/CO\,(SNe Ia + calibrated high-$z$ CO sample) are constructed to make comparisons for the constraint results. We could not obtain useful constraints with CO data alone because the CO data lack an important redshift range of $0.08<z<1.0$, where the SNe data are complimentary. Hence, we constrain cosmological parameters with the SNe and SNe/CO datasets to see the possible improvement for the SNe data by comparing the constraint contour.
   
\begin{figure*}
\centering
\begin{minipage}[t]{0.38\textwidth}
\centering
\includegraphics[scale=0.53]{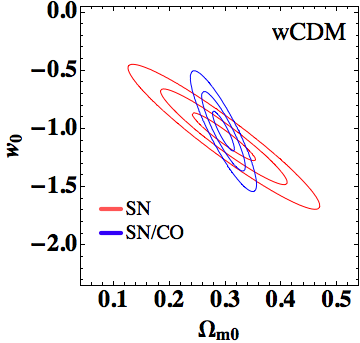}
\end{minipage}
\begin{minipage}[t]{0.38\textwidth}
\centering
 \includegraphics[scale=0.52]{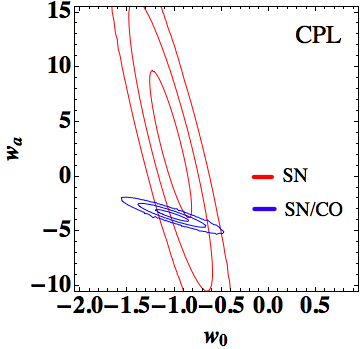}
\end{minipage}
\caption{The left panel is the $1-,2-,3-\sigma$ confidence levels for the datasets of the SNe\,(the red lines) and SNe/CO\,(the blue lines) on the constraint on the $w_{0}$ and $\Omega_{m0}$ of the $w$CDM model. The right panel is the $1-,2,-3\sigma$ confidence levels on the constraint of the $w_{0}$ and $w_{a}$ of the CPL model with the datasets of the SNe\,(the red contours) and SNe/CO\,(the blue contours).}
\label{fig: contour wCDM and CPL}
\end{figure*}

Fig.~\ref{fig: contour wCDM and CPL} presents the $1-,2-,3-\sigma$ confident levels of the two datasets, SNe\,(red lines) and SNe/CO\,(blue lines), for the $w$CDM model in the $w_{0}-{\Omega}_{m0}$ plane and CPL in the $w_{0}-w_{a}$ plane. Table~\ref{tab:constraint parameters} presents the constrained values of the parameters in the $w$CDM and CPL models. 

\begin{table*}
\caption{The constrained parameters in the $w$CDM and CPL models with the SNe and SNe/CO datasets. The values in the parentheses are the $1-\sigma$ uncertainties.}
\begin{tabular}[ht]{lcccc} 
\label{tab:constraint parameters} 
Fitting Dataset & $w_{0}$ & $w_{a}$ & $\Omega_{m0}$ & $\rm{H}_{0}\,(\rm{km\,s^{-1}\,Mpc^{-1}})$ \\
 \toprule[3pt]
 &&wCDM&&\\
 \hline
 &&&&\\
 SNe & $-1.07\pm{0.21}$ & $0$ & $0.30\pm{0.06}$ & $70.18\pm{0.42}$\\
 &&&&\\ 
 \hline
 &&&&\\
 SNe/CO & $-1.02\pm{0.17}$ & $0$ & $0.30\pm{0.02}$ & $70.00\pm{0.60}$\\
 &&&&\\
 \midrule[2pt]
  &&CPL&&\\
  \hline
 &&&&\\
 SNe & $-1.04\pm{0.22}$ & $2.96\pm{6.78}$ & $0.28\pm{0.11}$ & $70.18\pm{0.54}$\\
 &&&&\\
 \hline
 &&&&\\
 SNe/CO & $-1.03\pm{0.18}$ & $-3.54\pm{0.55}$ & $0.40\pm{0.06}$ & $70.00\pm{0.60}$\\
 &&&&\\
 \bottomrule[3pt]
\end{tabular}
\end{table*}
In our $w$CDM constraint, as shown in Table~\ref{tab:constraint parameters}, $\Omega_{m0}$ and $w_{0}$ in the SNe dataset are consistent with the results of $w$CDM constraint from the Table $7$ in \citet{2012ApJ...746...85S}\,\,($\Omega_{m0}=0.296^{+0.102}_{-0.180}$, $w_{0}=-1.001^{+0.348}_{-0.398}$, both with systematic uncertainties). We noted that the best-fit values of the $\Omega_{m0}$ and $w_{0}$ in the SNe dataset are consistent with these values in the SNe/CO set within their $1-\sigma$ uncertainties. The uncertainties of the $\Omega_{m0}$ and $w_{0}$ become smaller as the calibrated high-$z$ CO sample galaxies are added in the constraint. This case is also reflected on the SNe and SNe/CO contours in the left panel of Fig.~\ref{fig: contour wCDM and CPL}, that the shape of the contour shrinks after the joint of the calibrated high-$z$ CO sample.
     
In our CPL constraint, the goal is to investigate the evolution factor\,($w_{a}$) in Eq.~\ref{eq:wz} as considering the addition of the calibrated high-$z$ CO sample. In Table~\ref{tab:constraint parameters}, the $w_{0}$ and $w_{a}$ are both consistent between the SNe and SNe/CO datsets within a $1-\sigma$ level although the best-fit value of $w_{a}$ decreases after we included the calibrated high-$z$ CO galaxies. The uncertainties of the best-fit $w_{a}$ and $w_{0}$ in the SNe/CO dataset are smaller than these in the SNe dataset after the joint of the calibrated high-$z$ CO sample. As noted in the right panel of Fig.~\ref{fig: contour wCDM and CPL}, the shape of the contour shrinks significantly as the calibrated high-$z$ CO sample is involved in the constraint. The best-fit value of $\Omega_{m0}$ becomes increased and its uncertainty becomes small when the high-$z$ CO sample is included.     

In summary, in Fig.~\ref{fig:Hubble Diagram} and ~\ref{fig:muob-muth vs z}, some of the DMs of the CO data become lower beyond $z = 3$ as compared to the main concordance CDM\,(the gray line,\,($\Omega_{m0}$, $\Omega_{\Lambda 0}$, $w_{0}$, $w_{a}$)\,=\,($0.295$, $0.705$, $-1$, $0$)). We found that the addition of the calibrated high-$z$ CO data brings an improvement on the constraint of the $\Omega_{m0}$ and $w_{0}$ in the $w$CDM model. Also, the addition of the CO data shows a significant improvement on the constraint of the EoS of dark energy in the CPL model. Importantly, in the CPL constraint, the best-fit value of $w_{a}$ is over $1\sigma$ away from the null after adding the calibrated CO sample galaxies, indicating that the potential use of high-$z$ CO galaxies on tracing the evolution of dark energy. The observational errors will be reduced and more galaxy data will be provided in the near future with ALMA and other future surveys.
\section{Conclusions}
In this study, we explored the correlation between the CO($J=1-0$) luminosity\,($L^{\rm{intrins.}}_{\rm{CO}}$) and FWHM\,($W^{\rm{uncorr.}}_{\rm{CO}}$) of the CO-emitted galaxies spanning a large redshift range\,($0 < z < 7$). Our analysis shows : (i) a significant power-law relation between $L^{\rm{intrins.}}_{\rm{CO}}$ and $W^{\rm{uncorr.}}_{\rm{CO}}$ among the CO galaxies we selected at both $z < 1$ and $z > 1$, and (ii) no significant sign of redshift evolution of this correlation. 

Based on the result of no significant redshift evolution, we first calibrated the $L^{\rm{intrins.}}_{\rm{CO}}-W^{\rm{uncorr.}}_{\rm{CO}}$ relation with the nearby distance calibrators (e.g., SNe Ia and Cepheid-calibrated galaxies), and used this calibrated relation to obtain the cosmology-independent luminosity distances\,($D_{L}$) or DMs to our high-$z$ sample galaxies, and extended the Hubble Diagram into high-$z$ regions with our calibrated CO high-$z$ data. We then utilized these calibrated DMs of the high-$z$ sample galaxies to constrain cosmological parameters, with the SNe data. 

The parameters that were constrained in the $w$CDM model improved on $\Omega_{m0}$ and $w_{0}$ by adding the calibrated high-$z$ sample galaxies. Furthermore, the improvement on the constraint contour of the $w_{0}-w_{a}$ plane in the CPL model is significant. Through this preliminary test with the calibrated high-$z$ sample galaxies on the cosmological constraints, we present the potential application of CO data as an indicator to probe dark energy. Once the errors of individual CO measurements are reduced, tighter constraints can be obtained. Therefore, it is essential to increase the number of CO galaxies and improve the distance calibration so that constraints can become more reliable. We expect more galaxies to be observed in future surveys.

\section*{Acknowledgments}
We thank the referee for kind and useful suggestions, and thank for helping us to revise the paper. We thank the support from the Ministry of Science and Technology of Taiwan through grant NSC 103-2112-M-007-002-MY3, and NSC 105-2112-M-007-003-MY3, and the help from National Tsing Hua University in Taiwan. EKE acknowledges a post-doctoral fellowship from TUBITAK-BIDEB through 2218 program.
\bibliographystyle{mnras}
\bibliography{CO_low_high_z}
\appendix
\section{Galaxies in our samples} 
\label{S:Appendixgalaxy} 
\begin{table*}
\centering
\caption{Five IRAM galaxies were selected from \citet{1997ApJ...478..144S} and $53$ FCRAO galaxies from \citet{1995ApJS...98..219Y} into our low-$z$\,($z < 1$) sample. The first column is the ID of the galaxy. The second is the published redshift\,($z$) of each galaxy. The third is the velocity-integrated flux in $\rm{Jy\,km\,s^{-1}}$ with the $1-\sigma$ uncertainty. The fourth and fifth columns are the published velocity-integrated CO intensity\,($I_{CO}$, $\rm{K\,km\,s^{-1}}$) and peak antenna temperature\,($A$, $\rm{mK}$) and their corresponding $1-\sigma$ uncertainties for FCRAO sample galaxy in \citet{1995ApJS...98..219Y}. The sixth column is the published FWHM in $\rm{km\,s^{-1}}$. The last column is the reference. For the FCRAO galaxy, the velocity-integrated flux is a global-fitting flux. The FWHM measurement is given in the literature. The uncertainty of the FWHM measurement can be derived from the errors of the $I_{CO}$ and $A$, which is described in Sec.~\ref{low-z}. For the IRAM galaxy, the velocity-integrated flux listed in this table is converted by multiplying $4.5\rm{Jy\,K^{-1}}$ with the velocity-integrated CO intensity\,($I_{CO}$)\,\citep{1997ApJ...478..144S}. '$-$' denotes no value is given for the IRAM galaxy. The uncertainty of the FWHM measurement for the IRAM galaxy is $30\,\rm{km\,s^{-1}}$.}
\label{tab:tabb1}
\begin{tabular}{lllllll}
\toprule[3pt]
ID & $z$ & Flux\,($\rm{Jy\,km\,s^{-1}}$)& $I_{CO}$\,($\rm{K\,km\,s^{-1}}$)& $A$\,($\rm{mK}$) &FWHM\,($\rm{km\,s^{-1}}$)  &Reference \\
\midrule[2pt]
IZw36		&$0.000937$ 	&$20\pm{4}			$	&$0.45\pm{0.1}		$	&	$13\pm{3}	$  		&		$32$ 	&	\citet{1995ApJS...98..219Y}\\
NGC5195		&$0.001551$		&$240\pm{40}		$ 	&$3.57\pm{0.52}		$	&	$93\pm{8}	$		&		$40$    &	\citet{1995ApJS...98..219Y}\\
IIZw40		&$0.002632$ 	&$20\pm{4}			$	&$0.46\pm{0.1}		$	&	$14\pm{3}	$  		&		$42$  	&	\citet{1995ApJS...98..219Y}\\
NGC3949		&$0.002669$		&$220\pm{40}		$ 	&$3.87\pm{0.64}		$	&	$22\pm{3}	$		&		$150$   &	\citet{1995ApJS...98..219Y}\\
NGC6207		&$0.002842$		&$100\pm{20}		$ 	&$1.52\pm{0.45}		$	&	$17\pm{3}	$		&		$280$  	&	\citet{1995ApJS...98..219Y}\\
NGC4237		&$0.002892$		&$80\pm{15}			$ 	&$1.42\pm{0.3}		$	&	$19\pm{6}	$		&		$240$ 	&	\citet{1995ApJS...98..219Y}\\
NGC3353		&$0.003149$		&$20\pm{6}			$ 	&$0.55\pm{0.14}		$	&	$13\pm{3}	$		&		$41$ 	&	\citet{1995ApJS...98..219Y}\\
NGC4984		&$0.004266$		&$390\pm{70}		$ 	&$6.79\pm{0.79}		$	&	$39\pm{5}	$		&		$315$ 	&	\citet{1995ApJS...98..219Y}\\
NGC2964		&$0.004430$		&$340\pm{60}		$ 	&$6.25\pm{0.83}		$	&	$32\pm{4}	$		&		$280$ 	&	\citet{1995ApJS...98..219Y}\\
DDO218		&$0.004647$		&$100\pm{20}		$ 	&$1.81\pm{0.3}		$	&	$13\pm{3}	$		&		$130$  	&	\citet{1995ApJS...98..219Y}\\
NGC1022		&$0.004847$		&$310\pm{60}		$	&$5.32\pm{0.63}		$	&	$62\pm{5}	$		&		$110$  	&	\citet{1995ApJS...98..219Y}\\
NGC2748		&$0.004923$		&$160\pm{30}		$	&$3.00\pm{0.54}		$	&	$17\pm{5}	$		&		$350$	&	\citet{1995ApJS...98..219Y}\\
NGC7625		&$0.005447$ 	&$270\pm{50}		$	&$4.83\pm{0.74}		$	&	$37\pm{6}	$		&		$170$	&	\citet{1995ApJS...98..219Y}\\
NGC4039		&$0.005474$		&$920\pm{160}		$	&$16.32\pm{1.60}	$	&	$106\pm{10}	$		&		$230$	&	\citet{1995ApJS...98..219Y}\\
NGC2799		&$0.005581$		&$60\pm{10}			$	&$1.41\pm{0.32}		$	&	$29\pm{4}	$		&		$100$	&	\citet{1995ApJS...98..219Y}\\
NGC5861		&$0.006174$		&$350\pm{60}		$	&$5.59\pm{0.68}		$	&	$51\pm{3}	$		&		$180$	&	\citet{1995ApJS...98..219Y}\\
NGC5953		&$0.006555$		&$320\pm{60}		$	&$5.41\pm{0.69}		$	&	$44\pm{3}	$		&		$230$	&	\citet{1995ApJS...98..219Y}\\
NGC4385		&$0.007138$		&$70\pm{10}			$	&$1.28\pm{0.44}		$	&	$29\pm{3}	$		&		$60$	&	\citet{1995ApJS...98..219Y}\\
NGC4418		&$0.007268$		&$150\pm{30}		$	&$2.70\pm{0.54}		$	&	$26\pm{6}	$		&		$160$	&	\citet{1995ApJS...98..219Y}\\
NGC6574		&$0.007612$		&$680\pm{120}		$	&$12.30\pm{1.48}	$	&	$69\pm{6}	$		&		$350$	&	\citet{1995ApJS...98..219Y}\\
NGC4793		&$0.008286$		&$70\pm{15}			$	&$1.28\pm{0.36}		$	&	$22\pm{3}	$		&		$60	$	&	\citet{1995ApJS...98..219Y}\\
NGC4194		&$0.008342$		&$180\pm{30}		$	&$3.17\pm{0.4}		$	&	$31\pm{3}	$		&		$140$	&	\citet{1995ApJS...98..219Y}\\
NGC2750		&$0.008920$ 	&$140\pm{20}		$	&$2.31\pm{0.32}		$	&	$30\pm{3}	$		&		$96$	&	\citet{1995ApJS...98..219Y}\\
NGC7714		&$0.00933$		&$130\pm{20}		$	&$2.39\pm{0.44}		$	&	$20\pm{3}	$		&		$100$	&	\citet{1995ApJS...98..219Y}\\
NGC4433		&$0.010007$		&$270\pm{50}		$	&$5.20\pm{0.61}		$	&	$28\pm{3}	$		&		$310$	&	\citet{1995ApJS...98..219Y}\\
NGC3690		&$0.01041$		&$610\pm{110}		$	&$10.69\pm{1.15}	$	&	$45\pm{4}	$		&		$260$	&	\citet{1995ApJS...98..219Y}\\
NGC5653		&$0.01119$		&$280\pm{50}		$	&$5.01\pm{0.75}		$	&	$30\pm{3}	$		&		$270$	&	\citet{1995ApJS...98..219Y}\\
NGC877		&$0.01305$		&$430\pm{80}		$	&$7.08\pm{0.79}		$	&	$42\pm{3}	$		&		$220$	&	\citet{1995ApJS...98..219Y}\\
IR1510+07	&$0.0131$ 		&$150\pm{30}		$	&$3.42\pm{0.60}		$	&	$18\pm{2}	$		&		$290$	&	\citet{1995ApJS...98..219Y}\\
IC694		&$0.0132$		&$290\pm{50}		$	&$5.71\pm{0.65}		$	&	$35\pm{8}	$		&		$250$	&	\citet{1995ApJS...98..219Y}\\
NGC6701		&$0.01323$		&$340\pm{60}		$	&$5.87\pm{0.79}		$	&	$43\pm{3}	$		&		$140$	&	\citet{1995ApJS...98..219Y}\\
NGC5936		&$0.013356$ 	&$250\pm{40}		$	&$4.52\pm{0.64}		$	&	$34\pm{3}	$		&		$190$	&	\citet{1995ApJS...98..219Y}\\
NGC3221		&$0.01371$		&$400\pm{70}		$	&$6.29\pm{0.80}		$	&	$32\pm{2}	$		&		$150$	&	\citet{1995ApJS...98..219Y}\\
NGC7770		&$0.013733$		&$540\pm{100}		$	&$9.06\pm{1.08}		$	&	$53\pm{3}	$		&		$240$	&	\citet{1995ApJS...98..219Y}\\
NGC6921		&$0.01447$		&$350\pm{60}		$	&$7.44\pm{0.89}		$	&	$24\pm{4}	$		&		$740$	&	\citet{1995ApJS...98..219Y}\\
NGC23      	&$0.015231$		&$280\pm{50}		$	&$4.67\pm{0.54}		$	&	$16\pm{2}	$		&		$410$	&	\citet{1995ApJS...98..219Y}\\
NGC834		&$0.015321$		&$170\pm{30}		$	&$3.25\pm{0.67}		$	&	$21\pm{2}	$		&		$410$	&	\citet{1995ApJS...98..219Y}\\
NGC1614		&$0.01594$		&$290\pm{50}		$	&$5.75\pm{0.75}		$	&	$42\pm{4}	$		&		$250$	&	\citet{1995ApJS...98..219Y}\\
NGC2532		&$0.01752$		&$170\pm{30}		$	&$2.86\pm{0.50}		$	&	$32\pm{3}	$		&		$140$	&	\citet{1995ApJS...98..219Y}\\
NGC1275		&$0.01756$		&$50\pm{10}			$	&$0.85\pm{0.19}		$	&	$13\pm{3}	$		&		$70$	&	\citet{1995ApJS...98..219Y}\\
UGC2982		&$0.017696$		&$350\pm{60}		$	&$6.85\pm{0.87}		$	&	$35\pm{3}	$		&		$150$	&	\citet{1995ApJS...98..219Y}\\
NGC828		&$0.01793$		&$610\pm{110}		$	&$9.58\pm{1.10}		$	&	$36\pm{2}	$		&		$380$	&	\citet{1995ApJS...98..219Y}\\
Arp220		&$0.018126$ 	&$566\pm{100}		$	&$9.57\pm{0.53}		$	&	$30\pm{3}	$		&		$420$	&	\citet{1995ApJS...98..219Y}\\
NGC6286		&$0.01835$  	&$300\pm{50}		$	&$5.30\pm{0.65}		$	&	$14\pm{1}	$		&		$500$	&	\citet{1995ApJS...98..219Y}\\
Mrk331		&$0.018483$		&$450\pm{80}		$	&$9.17\pm{0.99}		$	&	$33\pm{2}	$		&		$320$	&	\citet{1995ApJS...98..219Y}\\
NGC2623		&$0.01851$		&$170\pm{30}		$	&$3.84\pm{0.56}		$	&	$19\pm{4}	$		&		$170$	&	\citet{1995ApJS...98..219Y}\\
NGC6240		&$0.02448$		&$300\pm{50}		$	&$5.68\pm{0.69}		$	&	$21\pm{2}	$		&		$420$	&	\citet{1995ApJS...98..219Y}\\
NGC5256		&$0.027863$ 	&$210\pm{40}		$	&$3.86\pm{0.83}		$	&	$23\pm{4}	$		&		$250$	&	\citet{1995ApJS...98..219Y}\\
NGC7674		&$0.02892$  	&$350\pm{60}		$	&$6.55\pm{1.11}		$	&	$61\pm{16}	$		&		$200$	&	\citet{1995ApJS...98..219Y}\\
NGC6090		&$0.029304$		&$200\pm{40}		$	&$3.46\pm{0.53}		$	&	$25\pm{3}	$		&		$120$	&	\citet{1995ApJS...98..219Y}\\
NGC695		&$0.03247$  	&$220\pm{40}		$	&$4.76\pm{0.58}		$	&	$18\pm{3}	$		&		$350$	&	\citet{1995ApJS...98..219Y}\\
Arp55		&$0.039300$ 	&$200\pm{30}		$	&$3.89\pm{0.57}		$	&	$18\pm{2}	$		&		$250$	&	\citet{1995ApJS...98..219Y}\\
VIIZw31 	&$0.053670$ 	&$190\pm{30}		$	&$4.24\pm{0.57}		$	&	$29\pm{4}	$		&		$260$	&	\citet{1995ApJS...98..219Y}\\
Arp193		&$0.023349$		&$162.0\pm{32.4}	$			& $36.0\pm{7.2}$ &		--								&		$410$	&	\citet{1997ApJ...478..144S}\\
Mrk273		&$0.037773$		&$85.5\pm{17.1}		$		& $19.0\pm{3.8}$&	    --										&		$300$	&	\citet{1997ApJ...478..144S}\\
09320+6134	&$0.039311$		&$70.2\pm{14.04}	$			& $15.6\pm{3.12}$ &		--								&		$350$	&	\citet{1997ApJ...478..144S}\\
Mrk231		&$0.042170$		&$99\pm{19.8}		$		& $22\pm{4.4}$ &		--									&		$200$	&	\citet{1997ApJ...478..144S}\\
IZw1		&$0.061142$		&$31.5\pm{6.3}		$			& $7.0\pm{1.4}$&		--								&		$410$	&	\citet{1997ApJ...478..144S}\\

\bottomrule[3pt]
\end{tabular}
\end{table*}

\begin{table*}
\centering
\caption{$48$ high-$z$\,($z > 1$) galaxies were selected as our high-$z$ sample. The first column is the ID of the galaxy. The second is the published redshift\,($z$) of the galaxy. The third is the velocity-integrated flux in $\rm{Jy\,km\,s^{-1}}$ with the $1-\sigma$ uncertainty. The fourth is the FWHM in $\rm{km\,s^{-1}}$ with the $1-\sigma$ uncertainty. The last column is the reference. The symbol $^{\dagger}$ denotes that the galaxy is observed with a gravitationally lensed effect and its magnification factor\,($\mu$) can be found in the reference.}
\label{tab:tabb2}
\begin{tabular}{lllll}
\toprule[3pt]
ID & $z$  & Flux\,($\rm{Jy\,km\,s^{-1}}$)&FWHM\,($\rm{km\,s^{-1}}$)  &Reference \\
\midrule[2pt]
BzK 4171							&$1.465$		&	$0.20\pm{0.05}	$&$430	\pm{190}	$&	\citet{2013MNRAS.433..498A} \\
BzK21000							&$1.521$		&	$0.13\pm{0.03}	$&$480	\pm{220}	$&	\citet{2013MNRAS.433..498A}\\
BzK16000							&$1.524$		&	$0.20\pm{0.06}	$&$217	\pm{80}	$&	\citet{2013MNRAS.433..498A}\\
SPT J045247-5018.6$^{\dagger}$	&$2.0078$	&	$0.96\pm{0.12}	$&$612	\pm{59}	$&	\citet{2016MNRAS.457.4406A}\\
B1938+666$^{\dagger}$			&$2.0592$	& 	$0.93\pm{0.11}	$&$654	\pm{71}	$&	\citet{2016ApJ...827...18S}\\
J115820.2-013753$^{\dagger}$		&$2.1911$	& 	$0.74\pm{0.12}	$&$260	\pm{30}	$&	\citet{2012ApJ...752..152H}\\
SMM J123549						&$2.2015$	& 	$0.32\pm{0.04}	$&$542	\pm{47}	$&	\citet{2011MNRAS.412.1913I}\\
J133649.9+291801	$^{\dagger}$		&$2.2024$	& 	$0.93\pm{0.12}	$&$210	\pm{20}	$&	\citet{2012ApJ...752..152H}\\
BX610							&$2.2105$	& 	$0.18\pm{0.04}	$&$240	\pm{70}	$&	\citet{2013MNRAS.433..498A}\\
IRAS F10214+4724	$^{\dagger}$		&$2.2856$	& 	$0.337\pm{0.045}	$&$184	\pm{29}	$&	\citet{2011ApJ...730..108R}\\
J134429.4+303036$^{\dagger}$		&$2.301$		& 	$2.749\pm{0.39}	$&$1140	\pm{130}	$&	\citet{2012ApJ...752..152H}\\
J090302.9-014127$^{\dagger}$		&$2.3051$	& 	$1.0	\pm{0.13}	$&$270	\pm{30}	$&	\citet{2012ApJ...752..152H}\\
HXMM01$^{\dagger}$				&$2.3079$	& 	$1.7	\pm{0.3	}	$&$840	\pm{160}	$&	\citet{2013Natur.498..338F}\\
SMM J213511-0102$^{\dagger}$		&$2.3259$	& 	$2.3	\pm{0.1	}	$&$290	\pm{30}	$&	\citet{2010Natur.464..733S}\\
HE1104-1805$^{\dagger}$			&$2.3220$	& 	$0.56\pm{0.09}	$&$372	\pm{49}	$&	\citet{2016ApJ...827...18S}\\
SMM J163650						&$2.3847$	& 	$0.34\pm{0.04}	$&$777	\pm{82}	$&	\citet{2011MNRAS.412.1913I}\\
J084933.4+021443	$^{\dagger}$		&$2.41$		& 	$1.04\pm{0.37}	$&$1180	\pm{320}	$&	\citet{2012ApJ...752..152H}\\
SMM J163658						&$2.4494$	& 	$0.37\pm{0.07}	$&$695	\pm{24}	$&	\citet{2011MNRAS.412.1913I}\\
J141351.9-000026$^{\dagger}$		&$2.4782$	& 	$1.47\pm{0.17}	$&$500	\pm{40}	$&	\citet{2012ApJ...752..152H}\\
SMM J123707-SW					&$2.4861$	& 	$0.12\pm{0.02}	$&$330	\pm{71}	$&	\citet{2011MNRAS.412.1913I}\\
SMM J123707-NE					&$2.4870$	& 	$0.09\pm{0.02}	$&$471	\pm{82}	$&	\citet{2011MNRAS.412.1913I}\\
SMM J04431+0210$^{\dagger}$		&$2.5086$	& 	$0.26\pm{0.05}	$&$415	\pm{62}	$&	\citet{2010ApJ...723.1139H}\\
SPT J012506-4723.7$^{\dagger}$	&$2.5146$	& 	$2.70\pm{0.22}	$&$428	\pm{27}	$&	\citet{2016MNRAS.457.4406A}\\
Cloverleaf$^{\dagger}$			&$2.5564$	& 	$1.378\pm{0.25}	$&$468	\pm{94}	$&	\citet{2011ApJ...730..108R}\\
SMM J14011+0252$^{\dagger}$		&$2.5653$	& 	$0.32\pm{0.04}	$&$151	\pm{19}	$&	\citet{2013ApJ...765....6S}\\
J113243.1-005108$^{\dagger}$		&$2.5778$	& 	$0.66\pm{0.19}	$&$380	\pm{90}	$&	\citet{2012ApJ...752..152H}\\
J091840.8+023047$^{\dagger}$		&$2.5811$	& 	$1.04\pm{0.26}	$&$680	\pm{140}	$&	\citet{2012ApJ...752..152H}\\
VCVJ1409+5628					&$2.5836$	& 	$0.27\pm{0.08}	$&$487	\pm{101}	$&	\citet{2016ApJ...827...18S}\\
SDP.130$^{\dagger}$				&$2.6256$	& 	$0.760\pm{0.120}	$&$360	\pm{40}	$&	\citet{2012ApJ...752..152H}\\
MS1512-cB58$^{\dagger}$			&$2.7265$	& 	$0.052\pm{0.013}	$&$174	\pm{43}	$&	\citet{2010ApJ...724L.153R}\\
SPT 213404-5013.2$^{\dagger}$		&$2.7788$	& 	$1.00\pm{0.18}	$&$469	\pm{180}	$&	\citet{2016MNRAS.457.4406A}\\
RX J0911+0551$^{\dagger}$		&$2.7961$	& 	$0.205\pm{0.029}	$&$111	\pm{19}	$&	\citet{2011ApJ...730..108R}\\
J04135+10277$^{\dagger}$		&$2.8421$	& 	$0.37\pm{0.07}	$&$765	\pm{222}	$&	\citet{2016ApJ...827...18S}\\
SMM J14009+0252$^{\dagger}$		&$2.9332$	& 	$0.31\pm{0.02}	$&$412	\pm{24}	$&	\citet{2010ApJ...723.1139H}\\
HLSW-01$^{\dagger}$				&$2.9574$	& 	$1.14\pm{0.11}	$&$350	\pm{55}	$&	\citet{2011ApJ...733L..12R}\\
J090311.6+003906$^{\dagger}$		&$3.042$		& 	$1.11\pm{0.25}	$&$435	\pm{54}	$&	\citet{2011ApJ...726L..22F}\\
Cosmic Eye$^{\dagger}$			&$3.074$		& 	$0.077\pm{0.013}	$&$190	\pm{24}	$&	\citet{2010ApJ...724L.153R}\\
J113526.3-014605$^{\dagger}$		&$3.1276$	& 	$0.35\pm{0.08}	$&$210	\pm{30}	$&	\citet{2012ApJ...752..152H}\\
MG 0751+2716	$^{\dagger}$			&$3.1984$	& 	$0.494\pm{0.105}	$&$290	\pm{62}	$&	\citet{2011ApJ...739L..32R}\\
J114637.9-001132	$^{\dagger}$		&$3.2592$	& 	$0.99\pm{0.16}	$&$680	\pm{80}	$&	\citet{2012ApJ...752..152H}\\
SMM J13120+4242					&$3.408$		& 	$0.42\pm{0.07}	$&$1040	\pm{190}	$&	\citet{2006ApJ...650..614H}\\
4C60.07n							&$3.791$		& 	$0.09\pm{0.01}	$&$165	\pm{24}	$&	\citet{doi:10.1146/annurev.astro.43.051804.102221}\\ 
APM 08279+5255$^{\dagger}$		&$3.911$		& 	$0.168\pm{0.015}	$&$556	\pm{55}	$&	\citet{2009ApJ...690..463R}\\
MM18423+5938	$^{\dagger}$			&$3.929605$	& 	$0.35\pm{0.05}	$&$161	\pm{30}	$&	\citet{2011ApJ...739L..30L}\\
PSS J2322+1944$^{\dagger}$		&$4.1192$	& 	$0.19\pm{0.08}	$&$200	\pm{70}	$&	\citet{2002ApJ...575..145C}\\
BR1202-0725						&$4.6932$	& 	$0.124\pm{0.012}	$&$329	\pm{36}	$&	\citet{2006ApJ...650..604R}\\
TN J0924							&$5.203$		& 	$0.087\pm{0.017}	$&$325	\pm{75}	$&	\citet{doi:10.1146/annurev.astro.43.051804.102221}\\
HFLS3$^{\dagger}$					&$6.3369$	& 	$0.074\pm{0.024}	$&$280	\pm{118}	$&	\citet{2013Natur.496..329R}\\
\bottomrule[3pt]
\end{tabular}
\end{table*}
\bsp	
\label{lastpage}
\end{document}